\documentclass[prl,showpacs,preprintnumbers,amsmath,aps,twocolumn]{revtex4}

\def\duzomniejsze{<\kern-.7mm<}
\def\duzowieksze{>\kern-.7mm>}

\def\textbf#1{{\bf #1}}
\def\be{\begin{equation}}
\def\ee{\end{equation}}
\def\bea{\begin{eqnarray}}
\def\eea{\end{eqnarray}}
\def\bse{\begin{subequations}}
\def\ese{\end{subequations}}
\newcommand{\bei}{\begin{itemize}}
\newcommand{\eei}{\end{itemize}}
\newcommand{\bee}{\begin{enumerate}}
\newcommand{\eee}{\end{enumerate}}

\def\>{\rangle}
\def\<{\langle}
\def\blacksquare{\vrule height 4pt width 3pt depth2pt}
\def\dt#1{{{\kern -.0mm\rm d}}#1\,}

\def\blacksquare{\vrule height 4pt width 4pt depth0pt}

\def\echv{G_{HV}}

\begin{document}

\title{Irreversibility for All Bound Entangled States}
\author{Dong Yang$^{(1)(2)}$}
\email{dyang@ustc.edu.cn}

\author{Micha\l{} Horodecki$^{(1)}$}
\email{fizmh@univ.gda.pl}

\author{Ryszard Horodecki$^{(1)}$}
\email{fizrh@univ.gda.pl}

\author{Barbara Synak-Radtke$^{(1)}$}
\email{dokbs@univ.gda.pl}

\affiliation{$^{(1)}$Institute of Theoretical Physics and
Astrophysics, University of Gda\'nsk, 80--952 Gda\'nsk, Poland\\
$^{(2)}$Hefei National Laboratory for Physical
Sciences at Microscale \& Department of Modern Physics,
University of Science and Technology of China, Hefei, Anhui 230026, Peoples Republic of China}

\date{\today}

\begin{abstract}
We derive a new inequality for entanglement for a mixed  four-partite state.
Employing  this inequality, we  present a one-shot lower bound for
entanglement cost and prove that  entanglement cost is strictly larger than
zero for any entangled state. We demonstrate that irreversibility occurs in the
process of formation for {\it all} non-distillable
entangled states. In this way we solve a long standing problem,
of how "real" is entanglement of bound entangled states.
Using the new inequality we also
prove the impossibility of local cloning of a known entangled state.
\end{abstract}

\pacs{03.67.Hk, 03.65.Ca, 03.65.Ud}

\maketitle
Irreversibility in physical processes is one of the most fundamental phenomena both
in classical mechanics and in quantum mechanics. In quantum information theory,
entanglement plays a crucial role. As is well known, entanglement cannot be created
from scratch and cannot be increased under local operations and
classical communication (LOCC). Under LOCC operations, one can only
change the form of entanglement. And there are two elementary processes
for entanglement manipulation. One is formation of entanglement and the
other is distillation of entanglement \cite{Bennett1}. Formation is the
process to create a generic entangled state $\rho_{AB}$ from a singlet
state. Entanglement cost $E_c(\rho)$ is the minimal number of singlets needed to prepare a bipartite state $\rho_{AB}$ by LOCC in asymptotic regime of many copies.
Distillation is the process to obtain nearly perfect singlets from many
identical copies of given state $\rho_{AB}$ by LOCC. Distillable
entanglement $E_d(\rho)$ is defined as the maximal number of singlets that can
be drawn from $\rho_{AB}$. Distillation is important because
singlet can be used for quantum teleportation \cite{Bennett2}, which is
the basic brick in quantum communication protocols. In some sense, formation
and distillation are dual to each other. Now the problem arises of whether
the two processes of formation and distillation are reversible or
irreversible? Reversibility means $E_c(\rho_{AB})=E_{d}(\rho_{AB})$ and
irreversibility means $E_c(\rho_{AB})>E_{d}(\rho_{AB})$  \cite{irrev-locc}. Of course, we
know $E_c(\rho_{AB})\ge E_{d}(\rho_{AB})$ by the non-increasing axiom
under LOCC. Here we want to separate '$\ge$' into '$=$' and '$>$'. It
is proved that pure states can be converted reversibly \cite{Bennett3,Popescu,Lo}.
In contrast, it is commonly believed that irreversibility occurs for
nontrivial mixed states. Perhaps this deep belief comes from the strongest
indication that there exist so-called bound entangled states from which no
entanglement can be distilled, but for which, in order to create a single copy,
entanglement is required \cite{Michal}.
This means that for bound entangled states  $E_d=0$
and so called entanglement of formation $E_f >0$. Entanglement of formation
 is defined as $E_f(\rho_{AB})=\min{\sum_ip_iE(\phi^i_{AB})}$, where
the minimum is over all pure ensembles $\{|\phi^i\>_{AB},p_i\}$
satisfying $\rho_{AB}=\sum_ip_i(|\phi^i\>\<\phi^i|)_{AB}$, and
entanglement for pure state $\phi_{AB}$ is $E(\phi_{AB})=S(tr_A(|\phi\>\<\phi|)_{AB})$, where $S(\rho)$ is von Neumann entropy defined as $S(\rho)=-tr\rho\log{\rho}$. $E_f$ is interpreted
as entanglement   needed  to create a single copy of state.
It was proved  that the regularized form of entanglement of formation is equal to entanglement cost
$E_c(\rho)=\lim_{n\to\infty}E_f(\rho^{\otimes n})/n$ \cite{Hayden}.
It is clear that if $E_f$  is additive, then immediately we
get $E_c=E_f>0$ and irreversibility for bound entangled states is quickly
solved. Unfortunately, although additivity of $E_f$ is a very desirable
property and is deeply believed, the proof escapes from us by far.
Without additivity of $E_f$ , it is difficult to rule out the
possibility that the amount of entanglement needed per copy
vanishes in the asymptotic limit, that is $E_f>0$ but $E_c=0$.
Although this seems unlikely, it has not been disproved.
In \cite{termo} it was shown that an additive quantity $S(A)-S(AB)$
is a lower bound for $E_f$. However if the quantity is positive
then $E_d$ is also nonzero \cite{DevetakWinter-hash}, hence the bound cannot be useful in the
case of bound entangled states.

Irreversibility is proved for some special classes of mixed states
\cite{Vidal,Vollbrecht,Henderson}. There are two approaches to prove
irreversibility. One approach is to find a new entanglement measure that lies
strictly between $E_d$ and $E_c$ for nontrivial mixed states. Note that any
entanglement measure should satisfy $E_d\le E\le E_c$ \cite{3H}.
Here 'strictly' means $E_d< E< E_c$ for nontrivial cases. There exist a few
entanglement measures \cite{Vedral,Christandl} that are not 'strict' measures. The other approach is to find a quantity that is a lower bound of $E_c$ and is nonzero for entangled state, but is unnecessarily a good measure. In this Letter, we follow the second approach. We find a new inequality of entanglement for a mixed four-partite state that can be employed to provide a {\it one-shot} lower bound for entanglement cost and the lower bound is strictly larger than zero for any entangled state. Irreversibility is immediately obtained for {\it all} non-distillable entangled states.

First we recall a measure for classical correlation of bipartite state $\rho_{AB}$ proposed in \cite{HV},
\bea
C_{\rightarrow}(\rho_{A:B})&=&\max_{A_{i}^{\dagger}A_{i}}S(\rho_B)-\sum_{i}p_{i}S(\rho^{i}_{B}),\nonumber\\
C_{\leftarrow}(\rho_{A:B})&=&\max_{B_{i}^{\dagger}B_{i}}S(\rho_A)-\sum_{i}p_{i}S(\rho^{i}_{A}),\nonumber
\eea
where $A_{i}^{\dagger}A_{i}$ is a POVM performed on subsystem $A$,
$\rho^{i}_{B}=tr_A(A_i\otimes I\rho_{AB}A_i^{\dagger}\otimes I)/p_i$
is remaining state of $B$ after obtaining the outcome $i$ on $A$, and
$p_i=tr_{AB}(A_i\otimes I\rho_{AB}A_i^{\dagger}\otimes I)$. In general,
$C_{\rightarrow}(\rho_{A:B})\neq C_{\leftarrow}(\rho_{A:B})$. We denote
$C(\rho_{A:B})=\max\{C_{\rightarrow}(\rho_{A:B}),C_{\leftarrow}(\rho_{A:B})\}$.
It is  proved that $C(\rho_{A:B})=0$ if and only if
$\rho_{AB}=\rho_A\otimes \rho_B$ \cite{HV}. Now we define a new quantity for $\rho_{AB}$
\bea
G_{\leftarrow}(\rho_{A:B})=\inf\sum_ip_iC_{\leftarrow}(\rho^i_{A:B}),\nonumber\\
G_{\rightarrow}(\rho_{A:B})=\inf\sum_ip_iC_{\rightarrow}(\rho^i_{A:B}),\nonumber\\
\echv(\rho_{A:B})=\inf\sum_ip_iC(\rho^i_{A:B}),\nonumber
\eea
where infimum is taken over $\{\rho^i_{AB},p_i\}$, generally a mixed
ensemble of realization of $\rho_{AB}$. The function $\echv$ is not an
entanglement measure, however it can be called "entanglement parameter",
as it satisfies the following property:

{\bf Theorem 1.} $\echv(\rho_{A:B})=0$ if and only if $\rho_{AB}$ is separable.

{\it Proof.} It is easy to prove 'if' part. For the 'only if' part, it is  
sufficient to prove that if $G_{\leftarrow}=0$, then $\rho_{AB}$ is separable.

In \cite{Terhal} it was argued that $C_{\leftarrow}(\rho_{A:B})$ is asymptotically continuous. The fastest argument comes from the duality relation between dual states. For a tripartite pure state $|\phi\>_{ABC}$,  $\rho_{AB}=tr_C|\phi\>\<\phi|$ is dual to $\rho_{AC}=tr_B|\phi\>\<\phi|$ and vice versa. The duality relation between dual states is \cite{Koashi}
\be
S(\rho_A)=E_f(\rho_{A:C})+C_{\leftarrow}(\rho_{A:B}).\label{se:duality}
\ee
Further notice the fact that if states $\rho_{AB}$ and $\sigma_{AB}$ are
close to each other, then there exist purifications $\phi_{ABC}$ and $\psi_{ABC}$ such that the dual states $\rho_{AC}$ and $\sigma_{AC}$ are close \cite{Close}.
Since it is known that $E_f$ is continuous \cite{Nielsen} and entropy is continuous,
then $C_{\leftarrow}$ is continuous too. By {\it Proposition 3} in {\it Appendix}, from continuity of $C_{\leftarrow}$ it follows that there exists an optimal decomposition $\{\rho^i_{AB},p_i\}$ realizing $G_{\leftarrow}$ and the decomposition contains $d^2+1$ elements at most where $d$ is the dimension of Hilbert space ${\cal H}_{AB}$. If the state is entangled, there must be a non-product state in decomposition, and of course $C_{\leftarrow}$ is nonzero on this state (because for a non-product state there always exists Alice's measurement that is correlated with Bob's system \cite{HV}). Thus $G_{\leftarrow}$ is nonzero for every entangled state.\\

More generally, mixed convex roof from any continuous function that vanishes only
on product states, gives a function that vanishes only on separable states.
If, in addition, the function does not increase under conditioning upon
local classical register, its convex roof is entanglement measure.
(We prove this, and explore the consequences elsewhere). However $C_\leftarrow(A:B)$
{\it can} increase under conditioning on Alice's side (a counterexample
can be found in \cite{DiVincenzo-locking}).
\\

{\bf Lemma 1.} For any four-partite pure state $|\phi\>_{AA'BB'}$, the following inequality of entanglement is satisfied
\be
E_f(\phi_{AA':BB'})\ge E_f(\rho_{A:B})+C(\rho_{A':B'}),\label{se:inqu}
\ee
where $\rho_{AB}=tr_{A'B'}\phi_{AA'BB'}$ and $\rho_{A'B'}=tr_{AB}\phi_{AA'BB'}$.\\

{\it Proof.} Apply the duality relation (\ref{se:duality}) to four-partite pure state $\phi_{AA'BB'}$ and regard $AA'$ as one part, then we get
\bse
\bea
S(\rho_{AA'})&=&E_f(\rho_{AA':B})+C_{\leftarrow}(\rho_{AA':B'})\\
&\ge&E_f(\rho_{A:B})+C_{\leftarrow}(\rho_{A':B'}),
\eea
\ese
where $\ge$ comes from the fact that both $E_f$ and $C_{\leftarrow}$ is non-increasing under local operations \cite{Bennett1,HV}.
Similarly we obtain
\be
S(\rho_{BB'})\ge E_f(\rho_{A:B})+C_{\rightarrow}(\rho_{A':B'}).
\ee
So we get inequality (\ref{se:inqu}) that completes the proof.\\

{\bf Proposition 1. (Main inequality)} For a mixed four-partite state $\rho_{AA'BB'}$,
\be
E_f(\rho_{AA':BB'})\ge E_f(\rho_{A:B})+\echv(\rho_{A':B'}),\label{se:main}
\ee
where $\rho_{AB}=tr_{A'B'}\rho_{AA'BB'}$ and $\rho_{A'B'}=tr_{AB}\rho_{AA'BB'}$.\\

{\it Proof.} Now we consider a mixed four-partite state $\rho_{AA'BB'}$. Suppose the
optimal realization of $E_f$ of $\rho_{AA'BB'}$ is $\{\phi^i_{AA'BB'},p_i\}$, then
we have

\bse
\bea
&&E_f(\rho_{AA':BB'})=\sum_ip_iS(\rho^i_{AA'})\nonumber\\
&\ge&\sum_ip_iE_f(\rho^i_{A:B})+\sum_ip_iC(\rho^i_{A':B'})\label{se:m1}\\
&\ge&E_f(\rho_{A:B})+\sum_ip_iC(\rho^i_{A':B'})\label{se:m2}\\
&\ge&E_f(\rho_{A:B})+\echv(\rho_{A:B})\label{se:m3},
\eea
\ese
where (\ref{se:m1}) comes from (\ref{se:inqu})and (\ref{se:m2})
 from the convexity of $E_f$, and (\ref{se:m3}) from the definition
of $\echv$. This ends the proof. \blacksquare

From {\it Proposition 1}, an immediate corollary is as follows:

{\bf Corollary 1.} For a four-partite state $\rho_{AA'BB'}$, if the reduced
state $\rho_{A'B'}$ is entangled, then $E_f(\rho_{AA'BB'})> E_f(\rho_{A:B})$.\\

We will now use the above corollary to solve the open problem in \cite{MSS}.
In \cite{MSS}, impossibility of cloning a known
entangled state under LOCC is reduced to whether $E(\rho_{AA'BB'})>E(\rho_{AB})$ for
some entanglement measure when $\rho_{A'B'}$ is entangled \cite{Buzek-local-cloning}. By
{\it Corollary 1}, we obtain no-go theorem for LOCC cloning:

{\bf Proposition 2.} It is impossible to clone a known entangled state by LOCC.
\\

Let us now pass to  the proof of the main result of the Letter.

{\bf Theorem 2.} For any entangled state $\rho_{AB}$,
$E_c(\rho_{A:B})\ge \echv(\rho_{A:B})>0$.\\

{\it Proof.} Entanglement cost is defined as the asymptotic cost of
singlets to prepare a bipartite mixed state and proved to
be $E_c(\rho)=\lim_{n\to\infty}E_f(\rho^{\otimes n})/n$ \cite{Hayden}.
Now consider $\rho^{\otimes n}$ and we have
\begin{eqnarray}
&&E_f(\rho^{\otimes n})=E_f(\rho^{\otimes n-1}\otimes \rho)\nonumber \geq
 E_f(\rho^{\otimes n-1})+\echv(\rho) \geq \\
&&\ge\cdots\ge E_f(\rho)+(n-1)\echv(\rho),
\end{eqnarray}
all inequalities $\ge$ come from (\ref{se:main}). Then
\be
{E_f(\rho^{\otimes n})\over n}\ge {(n-1)\over n}\echv(\rho).
\ee
Let $n\to \infty$ and we get $E_c(\rho)\ge \echv(\rho)>0$. This ends the proof.\blacksquare

It is notable that $\echv$ is a {\it one-shot} lower bound for $E_c$, an
asymptotic quantity. Recall that irreversibility
means $E_c(\rho_{AB}) > E_d(\rho_{AB})$ in the processes of formation and
distillation. Irreversibility is proved for some specific mixed
states \cite{Vidal,Henderson,Vollbrecht}. It is conjectured that irreversibility
occurs for nontrivial mixed states, especially for non-distillable entangled states.
It is well known that any PPT (positive partial transpose) entangled state is bound entangled \cite{Michal} and
it is conjectured that NPT (negative partial transpose) bound entangled states exist \cite{Dur,DiVincenzo}. Whatever
the case is, we have the strict inequality $E_c>0$ for entangled states. Therefore we
conclude that {\it irreversibility occurs for any non-distillable
entangled state.}\\

Summarizing, we present a new inequality of entanglement for a
mixed four-partite state. Based on this inequality, the asymptotic quantity, entanglement
cost is lower bounded by a one-shot quantity which is strictly larger than
zero for entangled state. So irreversibility occurs in asymptotic manipulations of
entanglement for all non-distillable entangled states that solves
the problem announced in the original paper on bound entanglement \cite{Michal}. Also the
new inequality
is employed to prove no-go theorem, saying that it is impossible
to clone a known entangled state by LOCC operations.\\

{\it Acknowledgement} We would like to thank Pawe\l{} Horodecki  and Marco Piani
for useful discussions. We also thank Ujjwal Sen for helpful comments. The work is supported by Polish Ministry of Scientific Research and Information Technology under the (solicited) grant no.~PBZ-MIN-008/P03/2003 and by EC grants RESQ, contract no.~IST-2001-37559 and QUPRODIS, contract no.~IST-2001-38877.

\appendix
\section{Appendix}
\label{theappendix}
{\bf Definition.} For any continuous function $f$ of state on Hilbert space $C^d$ one defines a {\bf mixed convex roof}  $\hat f$
\be
\hat f(\rho)=\inf \sum_i p_i f(\rho_i),
\ee
where the infimum is taken over all finite decompositions $\sum_i p_i \rho_i=\rho$.\\

{\bf Proposition 3.} The infimum is attained, and the optimal ensemble can be chosen to have $d^2+1$ elements.\\

{\it Proof.}  We use standard techniques from information theory \cite{Csiszar} (see also \cite{Uhlmann}).
 First, let us show that for any finite decomposition $\rho=\sum_{i=1}^np_i\rho_i$,
we can provide a decomposition $\rho=\sum_{i=1}^{d^2+1} q_i \sigma_i$ with $d^2+1$ elements, such that
\be
\sum_{i=1}^np_if(\rho_i)=\sum_{i=1}^{d^2+1} q_i f(\sigma_i).
\ee
To this, consider convex hull ${\cal A}$ of the set $\{ (\rho_i, f(\rho_i) )\}_{i=1}^n$.
 The point $x=(\sum_{i=1}^n p_i \rho_i,\sum_{i=1}^n p_i f(\rho_i))$ belongs ${\cal A}$.
The set ${\cal A}$ is a compact convex set, actually a polyhedron,
in $d^2$-dimensional real affine space (this comes from the fact that states belongs to the real $d^2$ dimensional
space of Hermitian operators and have unit trace). The set of extremal points is included
in the set $\{\rho_i,f(\rho_i)\}_{i=1}^{n}$.
Then from Caratheodory theorem it follows that $x$ can be written as a convex combination of
at most $d^2+1$ extremal points, i.e. $x=\sum_{i_j} q_{i_j} (\rho_{i_j},f(\rho_{i_j}))$
where $j=1,\ldots d^2+1$. Writing $q_{i_j}=q_j$, $\rho_{i_j}=\sigma_j$ we get
$\sum_{i=1}^n p_i \rho_i=\sum_{j=1}^{d^2+1} q_j \sigma_j$ and $\sum_{i=1}^np_i f(\rho_i)=\sum_{j=1}^{d^2+1}q_j f(\sigma_j)$.
 Thus we have found a decomposition that has $d^2+1$ elements, and returns the same value of average,
so that the infimum can be taken solely over such decompositions.
 Then from continuity of the function and compactness of the set of states it follows that the infimum is attained.



\begin{thebibliography}{99}




\bibitem{Bennett1}
C.H. Bennett, D.P. DiVincenzo, J.A. Smolin, W.K. Wootters, Phys. Rev. A {\bf 54}, 3824 (1996).

\bibitem{Bennett2}
C.H. Bennett, G. Brassard, C. Crepeau, R. Jozsa, A. Peres, and W.K. Wootters, Phys. Rev. Lett. {\bf 70}, 1895 (1993).

\bibitem{irrev-locc}
It should be emphasized that irreversibility of distillation-formation process is connected with the fact that the class of allowed operations is restricted to LOCC which plays the basic role in quantum communication theory. There are investigations towards lifting irreversibility by enlarging class of operations \cite{APE,thermo-ent2002}.

\bibitem{thermo-ent2002}
M. Horodecki, J. Oppenheim and R. Horodecki, Phys. Rev. Lett. {\bf 89}, 240403 (2002).

\bibitem{APE}
K. Audenaert, M. Plenio and J. Eisert, Phys. Rev. Lett. {\bf 90}, 027901 (2003).


\bibitem{Bennett3}
C.H. Bennett, H.J. Bernstein, S. Popescu, and B. Schumacher, Phys. Rev. A {\bf 53}, 2046 (1996).

\bibitem{Popescu}
S. Popescu and D. Rohrlich, Phys. Rev. A {\bf 56}, R3319 (1997).

\bibitem{Lo}
H.-K. Lo and S. Popescu, Phys. Rev. Lett. {\bf 83}, 1459 (1999).

\bibitem{Michal}
M. Horodecki, P. Horodecki, and R. Horodecki, Phys. Rev. Lett. {\bf 80}, 5239 (1998).


\bibitem{Hayden}
P.M. Hayden, M. Horodecki, and B.M. Terhal, J. Phys. A: Math. Gen. {\bf 34}(35), 6891 (2001).


\bibitem{termo}
P. Horodecki, R. Horodecki, and M. Horodecki, Acta Phys. Slov. {\bf 48}, 141 (1998).


\bibitem{DevetakWinter-hash}
I. Devetak and A. Winter, Proc. R. Soc. Lond. A, {\bf 461}, 207 (2005).


\bibitem{Vidal}
G. Vidal, W. D\"ur, J.I. Cirac, Phys. Rev. Lett. {\bf 89}, 027901 (2002); G. Vidal, J.I. Cirac, Phys. Rev. A {\bf 65}, 012323 (2002); G. Vidal, J.I. Cirac, Phys. Rev. Lett. {\bf 86}, 5803 (2001).

\bibitem{Vollbrecht}
K.G.H. Vollbrecht, R.F. Werner, M.M. Wolf, Phys. Rev. A {\bf 69}, 062304 (2004).

\bibitem{Henderson}
L. Henderson and V. Vedral, Phys. Rev. Lett. {\bf 84}, 2263 (2000).

\bibitem{3H}
M. Horodecki, P. Horodecki, and R. Horodecki, Phys. Rev. Lett. {\bf 84}, 2014 (2000).

\bibitem{Vedral}
V. Vedral, M.B. Plenio, M.A. Rippin, and P.L. Knight, Phys. Rev. Lett. {\bf 78}, 2275 (1997).

\bibitem{Christandl}
M. Christandl and A. Winter, J. Math. Phys. {\bf 45} (3), 829 (2004).

\bibitem{HV}
L. Henderson and V. Vedral J. Phys. A {\bf 34}, 6899 (2001).

\bibitem{Terhal}
B.M. Terhal, M. Horodecki, D.P. DivVincenzo, and D. Leung, J. Math. Phys. {\bf 43}, 4286 (2002).

\bibitem{Koashi}
M. Koashi and A. Winter, Phys. Rev. A {\bf 69}, 022309 (2004).

\bibitem{Close}
We use trace-norm to measure hwo close two mixed states are and suppose $tr|\rho_{AB}-\sigma_{AB}|\le\epsilon$. Fidelity between two mixed states is defined as $F(\rho,\sigma)=tr(\sqrt{\sqrt{\rho}\sigma\sqrt{\rho}})$, and it is proved that $F(\rho,\sigma)=\max_{\phi,\psi}|\<\phi|\psi\>|$, where $\phi$ and $\psi$ are purifications of $\rho$ and $\sigma$ respectively \cite{Jozsa}. Fidelity and trace-norm satisfies the inequality \cite{Fuchs}
\be
1-F(\rho,\sigma)\le {tr|\rho-\sigma|\over 2}\le\sqrt{1-F^{2}(\rho,\sigma)}.\nonumber
\ee
Applying these formulas to $\rho_{AB}$ and $\sigma_{AB}$, we know that there exist purifications $\phi_{ABC}$ and $\psi_{ABC}$ such that
\be
|\<\phi_{ABC}|\psi_{ABC}\>|=F(\rho_{AB},\sigma_{AB})\ge 1-{\epsilon\over 2}.\nonumber
\ee
Then we get
\bea
F(\rho_{AC},\sigma_{AC})\ge|\<\phi_{ABC}|\psi_{ABC}\>|\ge 1-{\epsilon\over 2},\nonumber\\
tr|\rho_{AC}-\sigma_{AC}|\le\sqrt{\epsilon(4-\epsilon)}\le 2\sqrt{\epsilon}.\nonumber
\eea So $\rho_{AC}$ is close to $\sigma_{AC}$ when $\epsilon\to 0$.



\bibitem{Jozsa}
R. Jozsa, J. Mod. opt. {\bf 41}, 2315 (1994).

\bibitem{Fuchs}
C.A. Fuchs, J. Graaf, quant-ph/9712042.

\bibitem{Nielsen}
M.A. Nielsen, Phys. Rev. A {\bf 61}, 064301 (2000).

\bibitem{DiVincenzo-locking}
D. DiVincenzo, M. Horodecki, D. Leung, J. Smolin and B. Terhal, Phys. Rev. Lett. {\bf 92}, 067902 (2004).

\bibitem{MSS}
M. Horodecki, A. Sen De, U. Sen, Phys. Rev. A {\bf 70}, 052326 (2004).

\bibitem{Buzek-local-cloning}
V. Buzek, V. Vedral, M.B. Plenio, P.L. Knight, and  M. Hillery, Phys. Rev. A {\bf 55}, 3327 (1997).

\bibitem{Dur}
W. D\"ur, J.I. Cirac, M. Lewenstein, and D. Bruss, Phys. Rev. A {\bf 61}, 062313 (2000).

\bibitem{DiVincenzo}
D.P. DiVincenzo, P.W. Shor, J.A. Smolin, B.M. Terhal, and A.V. Thapliyal, Phys. Rev. A {\bf 61}, 062312
(2000).

\bibitem{Csiszar}
I. Csiszar and J. Korner, IEEE Trans. Inf. Theroy {\bf 24}, 339 (1978).

\bibitem{Uhlmann}
A. Uhlmann, Open Sys. Inf. Dyn. {\bf 5}, 209 (1998).



\end{thebibliography}
\end{document}